\documentclass[showpacs,amsmath,amssymb,twocolumn,prl,superscriptaddress]{revtex4-2}
\usepackage{amssymb}
\usepackage[dvips]{graphicx}
\usepackage{enumerate}
\usepackage{epsfig} 
\usepackage{subfigure}
\usepackage{xcolor}
\usepackage[T1]{fontenc}
\usepackage{fullpage}
\usepackage{amsthm,amsfonts,amssymb,amscd,mathrsfs,xspace,framed}
\usepackage{mathrsfs,amsmath}
\usepackage{color}
\usepackage{setspace}
\usepackage{url}
\usepackage{wrapfig}
\usepackage{tikz}
\usepackage{enumitem}
\usepackage{bm}
\usepackage{comment}
\usepackage{txfonts}
\usepackage{array}
\usepackage{color}
\usepackage{braket}
\usepackage{hyperref}
\usepackage[english]{babel}
\usepackage{url}
\usepackage{graphicx}
\usepackage{pdfpages}

\def\be{\begin{equation}}
\def\ee{\end{equation}}
\def\ba{\begin{eqnarray}}
\def\ea{\end{eqnarray}}

\makeatletter
\AtBeginDocument{\let\LS@rot\@undefined}
\makeatother

\def\be{\begin{equation}}
\def\ee{\end{equation}}
\def\ba{\begin{eqnarray}}
\def\ea{\end{eqnarray}}

\begin{document}

\date{\today}

\title{Entanglement-Enhanced Optomechanical Sensing}
\author{Yi Xia}
\email{yixia.light@gmail.com}
\affiliation{James C. Wyant College of Optical Sciences, University of Arizona, Tucson, Arizona 85721, USA}
\affiliation{Department of Materials Science and Engineering, University of Arizona, Tucson, Arizona 85721, USA}
\author{Aman R. Agrawal}
\affiliation{James C. Wyant College of Optical Sciences, University of Arizona, Tucson, Arizona 85721, USA}
\author{Christian M. Pluchar}
\affiliation{James C. Wyant College of Optical Sciences, University of Arizona, Tucson, Arizona 85721, USA}
\author{Anthony J. Brady}
\affiliation{Department of Electrical and Computer Engineering, University of Arizona, Tucson, Arizona 85721, USA}
\author{Zhen Liu}
\affiliation{School of Physics and Astronomy, University of Minnesota, Minneapolis, MN 55455, USA}
\author{Quntao Zhuang}
\affiliation{Department of Electrical and Computer Engineering, University of Arizona, Tucson, Arizona 85721, USA}
\affiliation{James C. Wyant College of Optical Sciences, University of Arizona, Tucson, Arizona 85721, USA}
\affiliation{Department of Electrical and Computer Engineering, University of Southern California, Los Angeles, CA 90089, USA}
\author{Dalziel J. Wilson}
\affiliation{James C. Wyant College of Optical Sciences, University of Arizona, Tucson, Arizona 85721, USA}
\author{Zheshen Zhang}
\email{zszh@umich.edu}
\affiliation{Department of Materials Science and Engineering, University of Arizona, Tucson, Arizona 85721, USA}
\affiliation{Department of Electrical and Computer Engineering, University of Arizona, Tucson, Arizona 85721, USA}
\affiliation{James C. Wyant College of Optical Sciences, University of Arizona, Tucson, Arizona 85721, USA}
\affiliation{Department of Electrical Engineering and Computer Science, University of Michigan, Ann Arbor, MI 48109, USA}

\begin{abstract}
Optomechanical systems have been exploited in ultrasensitive measurements of force, acceleration, and magnetic fields. The fundamental limits for optomechanical sensing have been extensively studied and now well understood---the intrinsic uncertainties of the bosonic optical and mechanical modes, together with the backaction noise arising from the interactions between the two, dictate the Standard Quantum Limit (SQL). %for the measurement sensitivity achievable by individual optomechanical sensors.
Advanced techniques based on nonclassical probes, in-situ pondermotive squeezed light, and backaction-evading measurements have been developed to overcome the SQL for individual optomechanical sensors. An alternative, conceptually simpler approach to enhance optomechanical sensing rests upon joint measurements taken by multiple sensors. In this configuration, a pathway toward overcoming the fundamental limits in joint measurements has not been explored. Here, we demonstrate that joint force measurements taken with entangled probes on multiple optomechanical sensors can improve the bandwidth in the thermal-noise-dominant regime or the sensitivity in shot-noise-dominant regime. Moreover, we quantify the overall performance of entangled probes with the sensitivity-bandwidth product and observe a $25\%$ increase compared to that of the classical probes. The demonstrated entanglement-enhanced optomechanical sensing could enable new capabilities for inertial navigation, acoustic imaging, and searches for new physics.

\end{abstract}


\maketitle

\section{\label{sec:level1}Introduction}

Optomechanical sensors~\cite{li2021cavity,liu2021progress} have garnered significant interest owing to their high sensitivity in measurements of force~\cite{gavartin2012hybrid}, acceleration~\cite{krause2012high}, and magnetic fields~\cite{forstner2012cavity}, immunity to electromagnetic interference, and small footprint~\cite{gavartin2012hybrid,krause2012high}. As extensively studied in the field of cavity optomechanics~\cite{aspelmeyer2014cavity}, the superior performance of optomechanical sensors stems from their low-noise readout mechanism based on parametric coupling of an optical field and a mechanical oscillator, in contrast to micro-electro-mechanical systems which are often plagued by technical noise. In cavity optomechanical sensors, a probe field is coupled into an optical cavity where a mechanical oscillator resides. Physical displacement of the mechanical oscillator shifts the cavity resonant frequency, which in turn shifts the phase of the field leaving the cavity. The sensitivity of the displacement measurement is typically bound by the Standard Quantum Limit (SQL) dictated by several fundamental noise sources including imprecision noise, also known as the shot noise owing to the photon-number fluctuations in the probe, and  backaction noise arising from the interaction between the radiation-pressure shot noise and the mechanical oscillator~\cite{caves1980measurement,clerk2010introduction,teufel2009nanomechanical}. Several techniques have been developed in recent years to improve sensitivity for individual optomechanical sensors. To combat imprecision noise, probes carrying squeezed light have been employed in Advanced LIGO to enable a 3-dB sensitivity improvement in the ongoing observation run~\cite{tse2019quantum}; and in optomechanical magnetometry, to enhanced the sensitivity and bandwidth in detecting magnetic fields~\cite{li2018quantum}. 
\begin{figure*}[bth!]
    \centering
    \includegraphics[width=0.95\textwidth]{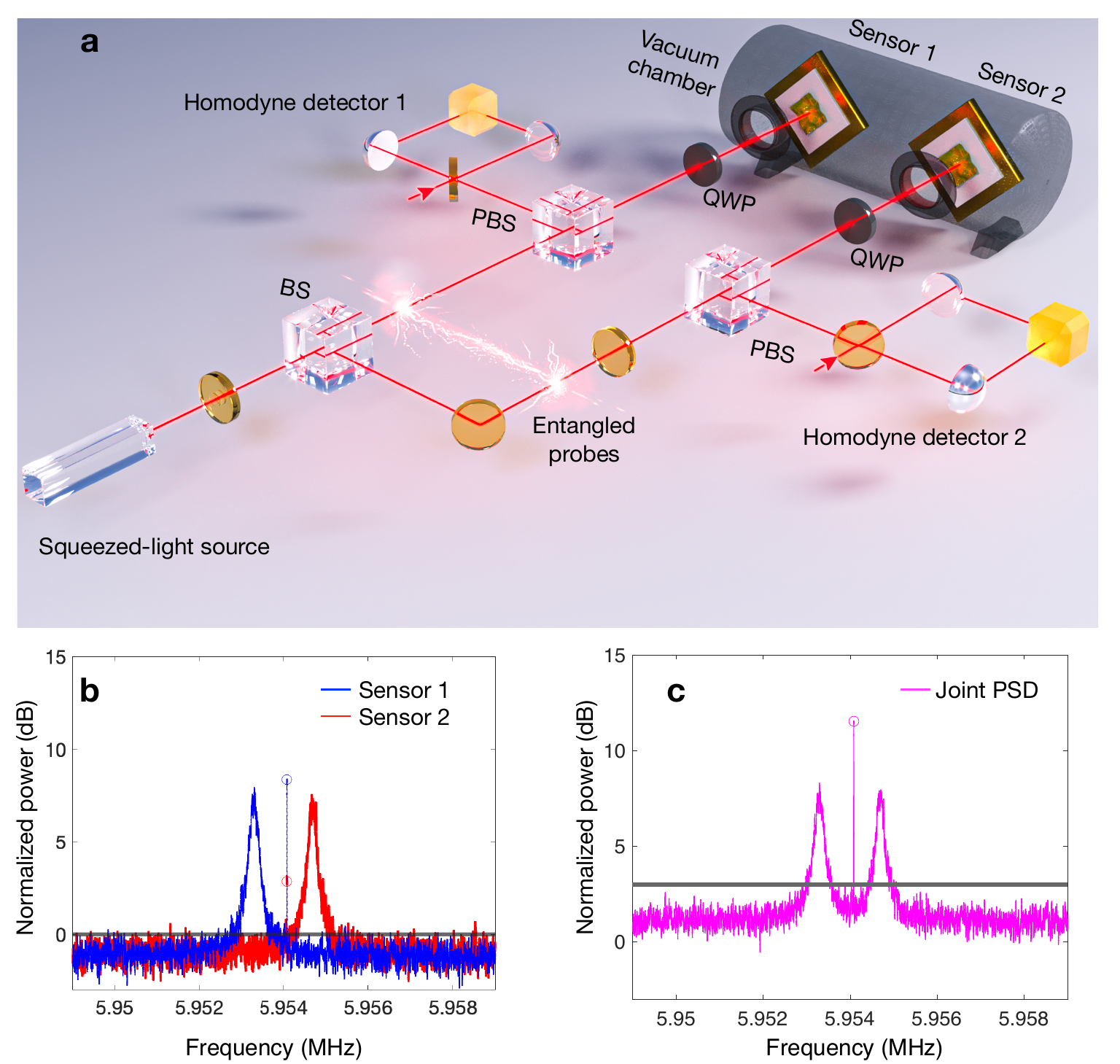}
    \caption{(a)Experiment setup. BS: beam splitter. PBS: polarizing beam splitter. QWP: quarter wave plate. (b) Normalized power spectral densities (PSDs) of individual homodyne measurement for the displacement of each membrane using entangled probes. Individual shot noise PSD (gray line) is normalized to unity. (c) Normalized PSD of joint homodyne measurement for the displacements of both membranes. Joint shot noise PSD (gray line) is normalized to 3 dB. Circles mark the delta peaks. }
    \label{fig:concept}
\end{figure*}
On the other hand, backaction-evading measurements, quantum non-demolition measurements, and imprecision-backaction correlations provide a route to beating the SQL. These approaches have been implemented in cavity optomechanics using two-tone driving~\cite{shomroni2019optical,ockeloen2016quantum}, negative mass oscillators~\cite{tsang2012evading, mercier2021quantum, moller2017quantum}, and the intrinsic optomechanical Kerr non-linearity~\cite{mason2019continuous,sudhir2017quantum,purdy2017quantum,kampel2017improving}.

Apart from these intriguing advances in optomechanical measurement techniques with a single sensor, a parallel %, albeit deemed scientifically less exciting, 
route to enhance optomechanical measurements builds on increasing the number of sensors. Per the central limit theorem, averaging the measurement outcomes of $M$ identical and independent sensors reduces the statistical uncertainty by a factor of $1/\sqrt{M}$. As such, a large number of sensors can boost the measurement sensitivity in detecting a common signal, a scenario pertinent to a wide range of sensing tasks from earthquake warning systems~\cite{d2019review} to dark matter searches~\cite{carney2021ultralight,PRXQuantum.3.030333,derevianko2018detecting,carney2021mechanical,carney2020proposal}.

Quantum metrology harnesses nonclassical resources to outperform the $1/\sqrt{M}$ factor of joint measurements, also known as the SQL scaling \cite{giovannetti2006quantum}. Distributed quantum sensing is a quantum-metrology paradigm that leverages entanglement shared by multiple sensors to achieve, in an ideal situation, a more favorable Heisenberg scaling of $1/M$ for the joint-measurement sensitivity~\cite{proctor2018multiparameter,zhang2021distributed,zhuang2018distributed,ge2018distributed}. Recent distributed quantum sensing experiments have demonstrated that entangled sensors outperform separable sensors in estimating global parameters such as the average optical phase shifts~\cite{guo2020distributed,liu2021distributed,hong2021quantum} and RF phase gradients~\cite{xia2020demonstration}. To date, entanglement-enhanced optomechanical sensing has not been explored. In this work, we make a critical step toward surpassing the SQL scaling for arrayed optomechanical sensors, by verifying that entangled probes improve joint force measurements with two mechanical membranes. We observe that entangled probes reduce the joint noise floor by 2 dB, leading to a 40\% improvement in the force sensitivity in the shot-noise-dominant regime. In addition, entangled probes also extend the frequency range over which thermal noise is dominant, thus enhancing the sensor bandwidth by 20\%. We further quantify the joint sensitivity and bandwidth with respect to resonant frequency difference. We assess the overall performance of joint force detection using the sensitivity-bandwidth product as a figure of merit.
Finally, we investigate joint force sensing of two incoherent forces, demonstrating that entangled probes can shorten the integration time by $60\%$ (limited by the 2 dB squeezing in the imprecision noise limit) and improve sensing bandwidth by $20\%$ in the thermal noise limit, accelerating spectral scanning rate in search of unknown signals.

\begin{figure*}[tbh!]
\vspace{-5pt}
    \centering
    \includegraphics[width=1\textwidth]{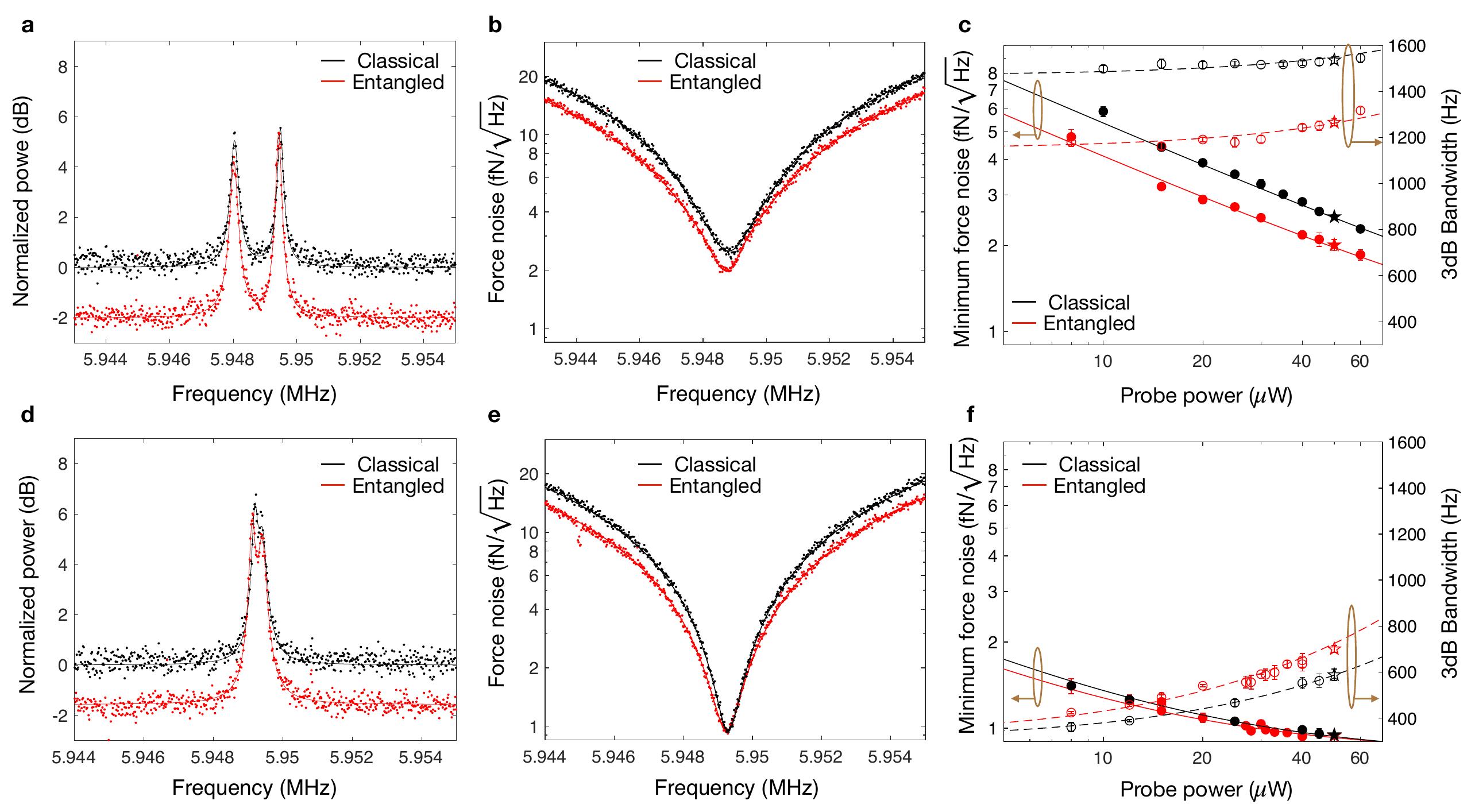}
    \vspace{-10pt}
    \caption{Entanglement-enhanced versus classical optomechanical sensing. (a, d) Normalized joint PSDs of homodyne measurements for displacements. Joint shot noise is normalized to unity. (b, e) Joint force noise at 50 $\mu$W probe power. (c, f)  Joint minimum force noise (solid lines) and bandwidth (dashed lines) at different probe power. Filled dots: minimum force noise; open dots: bandwidth; stars: fitted from force noise in (b, e). Resonant frequency difference: 1422 Hz in (a-c) and 262 Hz in (d-f). In all figures, entangled probes: red; classical probes: black; dots: experimental data; curves: theory. Error bars account for the frequency difference drifting, theory fitting, and fluctuations in the noise power measurements.}
    \label{fig:entangled_vs_classical}
\end{figure*}

\section{\label{sec:experiment}Experiment}

The workhorse for entanglement-enhanced optomechanical sensing is the squeezed light that is split into multiple arms to create entangled probes. The quantum advantage of this approach over separable classical probes stems from the correlated shot noise across the entangled probes, in the same vein as recent entangled sensor network experiments~\cite{guo2020distributed,xia2020demonstration}. Figure~\ref{fig:concept} (a) sketches the experimental setup. The probes couple to two separate optomechanical sensors each comprising a $100\times 100$ $\mu$m$^2$ Si$_3$N$_4$ membrane with a reflectively $R \approx 11.5\%$ atop a high reflectively ($R>99.9\%$) mirror, forming an optical cavity with a finesse $\sim 3$. Each membrane supports a set of high-Q drum modes with an effective mass of $6.75\times 10^{-13}$ kg at resonant frequencies of a few megahertz. We study sensing with the first higher-order mode of the two membranes at $\Omega_1/2\pi \sim 5.953$ MHz and $\Omega_2/2\pi \sim 5.955$ MHz, with damping rates of $\Gamma_1/2\pi \sim 200$ Hz and $\Gamma_2/2\pi \sim 260$ Hz. Homodyne measurements of the phase quadratures of the output probes from each cavity yield spectral amplitudes (SA) of
\begin{equation}
\hat{Y}_{\rm out}^{(i)}(\omega)=\hat{Y}_{\rm in}^{(i)}(\omega)+\alpha_i\beta_i\chi_i(\omega)\left[\hat{F}_{\rm th}^{(i)}(\omega)+\hat{F}_{\rm sig}^{(i)}(\omega)\right],
\end{equation}
where $i \in \{1,2\}$ is the sensor index, $\hat{Y}_{\rm in}^{(i)}(\omega)$ is the phase quadrature of the input probe, $\alpha_i^2$ is the mean photon number of each input probe and we assume $\alpha_i$ to be real for simplicity, $\beta_i=4\sqrt{2}G_i/\kappa_i$ is the optomechanical transduction efficiency. $G_i$ is the parametrical coupling between the cavity resonance frequency and mechanical oscillator position. $\kappa_i$ is the cavity decay rate. 
$\chi_i(\omega)=\frac{1/m_{\rm eff}}{\Omega_i^2-\omega^2+i\omega\Gamma_i}$ is the mechanical susceptibility. $m_{\rm eff}$ is the effective mass. $\hat{F}_{\rm th}^{(i)}(\omega)$ is the SA of the thermal force, and $\hat{F}_{\rm sig}^{(i)}(\omega)$ is the SA of the force signal. The estimation of the average force at two sensors of nearly equal optomechanical transduction efficiencies ($\beta_1\approx \beta_2=\beta$) is carried out using near-optimal entangled probes generated by evenly splitting the squeezed light into two arms ~\cite{guo2020distributed,xia2020demonstration} ($\alpha_1=\alpha_2=\alpha_c/\sqrt{2}$) where $\alpha_c^2$ is the mean photon number at carrier wavelength of squeezed light. To achieve optimal performance, a frequency-dependent entangled light needs to be engineered according to the force transduction efficiencies at each sensor over the entire sensing bandwidth (see Supplemental Material).

\section{\label{sec:results} Results}

To capture the physics behind entanglement-enhanced optomechanical sensing, we plot the normalized power spectral densities (PSDs), $S_{Y_{\rm out}^{(1)}Y_{\rm out}^{(1)}}(\omega)$ (blue) and $S_{Y_{\rm out}^{(2)}Y_{\rm out}^{(2)}}(\omega)$ (red), of the homodyne measurements for membrane displacements in Fig.~\ref{fig:concept} (b), in which the shot-noise level (SNL) is normalized to unity represented by the gray line. The overall detection efficiency at each sensor is $74\%$ (see Supplementary Information), leading to an imprecision noise floor $\sim$ 1 dB below the SNL, whereas the measured squeezing level from the source is $\sim$ 4 dB below the SNL~\cite{xia2020demonstration}. The spectra also show a thermal-noise-dominant band in the vicinity of the mechanical resonant frequencies, manifested as two broad peaks. Radiation pressure test forces on the membranes are created by an auxiliary amplitude-modulated 775-nm laser, yielding two delta peaks that are 2.8 dB and 8.36 dB above the SNL, highlighted by circles in Fig.~1(b). The signal-to-noise ratio (SNR) at each sensor is slightly improved due to the residue single-mode squeezing in each probe. Figure~1 (c) draws the joint homodyne PSD $S_{Y_{\rm out}^{\rm (joint)}Y_{\rm out}^{\rm (joint)}}$ obtained by adding the homodyne measurement records from both sensors, showing a more substantial SNR advantage for entangled probes. The signals coherently add to 11.5 dB, while the joint imprecision noise floor increases to 1 dB for entangled probes, as compared to the anticipated 3 dB for classical probes. 
Notably, this 2 dB noise reduction implies a joint sensitivity improvement beyond the $1/\sqrt{M}$ SQL scaling---of benefit for broadband, shot-noise-limited distributed force sensing applications, such as accelerometer arrays.

\subsection{Entanglement-enhanced measurement sensitivity}
\begin{figure*}[th]
    \centering
    \includegraphics[width=0.98\textwidth]{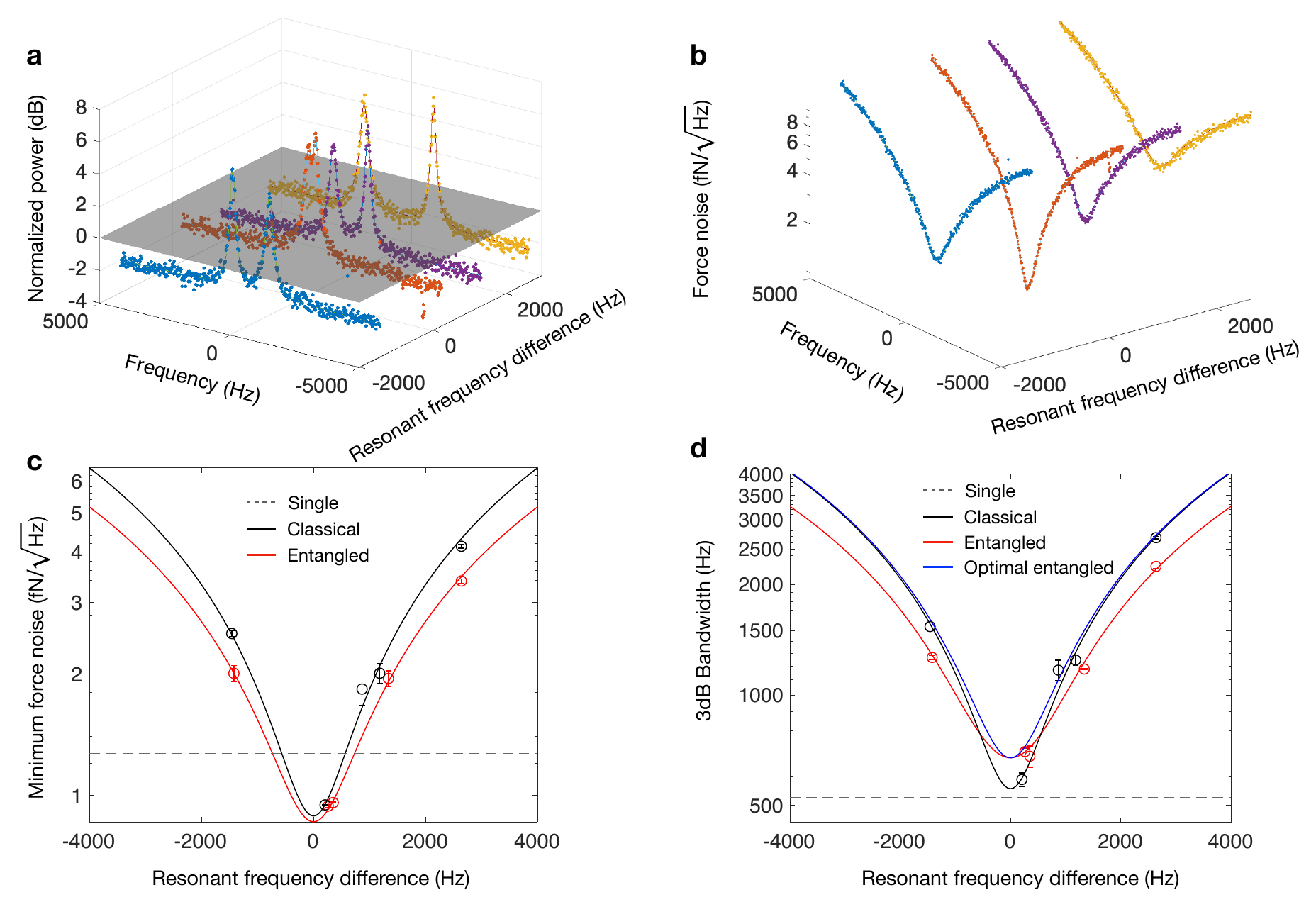}
    \vspace{-10pt}
    \caption{Sensitivity and bandwidth reconfigured by resonant frequency differences. (a) Normalized PSD of joint homodyne measurement and (b) joint force sensitivities based on entangled probes at 50 $\mu$W power. In (a, b), blue, red, purple, and yellow curves correspond to resonant frequency differences of -1422, 262, 1339, and 2641 Hz. Shaded plane: shot-noise level. (c) Sensitivity and (d) bandwidth at various resonant frequency differences for entangled (red), classical (black), and optimal entangled (blue) probes with 50 $\mu$W power. Circles: experimental data; solid lines: theory; dashed lines: single classical sensor. Error bars account for frequency difference drifting, theory fitting, and fluctuations in the noise power measurements.}
    \label{fig:frequency_difference}
\end{figure*}

We next investigate the performance of averaged force $\bar{F}=(F^{(1)}+F^{(2)})/2$ estimation based on entangled or classical probes interrogating two mechanical sensors with a large resonant frequency difference (1422 Hz). Figure~\ref{fig:entangled_vs_classical} (a) shows the joint homodyne PSDs for the two cases with the joint SNL normalized to unity. The noise peaks in the vicinity of the mechanical resonant frequencies are of equal magnitude for the entangled and classical probes due to the dominant thermal noise in this region. Nonetheless, the entangled probes reduce the off-resonant imprecision noise floor from the SNL by 2 dB. To characterize the advantage this offers for force sensing, Fig.~\ref{fig:entangled_vs_classical} (b) shows the joint force noise $\sqrt{S_{\bar{F}\bar{F}}}$ derived by rescaling the individual output phase quadratures: $\hat{F}^{(i)}=\hat{Y}_{\rm out}^{(i)}(\omega)/(\alpha_i\beta_i |\chi_i(\omega)|)$ and adding them together(see Supplementary Information). The minimum force noise PSDs $S_{\bar{F}_{\rm min}\bar{F}_{\rm min}}$ associated with the classical probes (black) is achieved at $\omega_{\rm min}=\frac{1}{2}\sqrt{-\Gamma_1^2-\Gamma_2^2+2(\Omega_1^2+\Omega_2^2)}\approx \frac{\Omega_1+\Omega_2}{2} \equiv \bar{\Omega}$ in the shot-noise-dominant region between the two resonant frequencies. The two mechanical susceptibilities coincide $|\chi_1(\bar{\Omega})|=|\chi_2(\bar{\Omega})|$ near $\bar{\Omega}$. The entangled probes (red) reduce the force noise by $20\%$. In Fig.~\ref{fig:entangled_vs_classical} (c), we plot the minimum force noise $\sqrt{S_{\bar{F}_{\rm min}\bar{F}_{\rm min}}}$ (filled dots) for the entangled (red) and classical (black) probes at different power levels and solid lines are theoretical predictions. The minimum force noise PSD is approximated by 
\begin{equation}
S_{\bar{F}_{\rm min}\bar{F}_{\rm min}}\approx\frac{m^2_{\rm eff}}{\beta^2\alpha_c^2}\bar{\Omega}^2(\bar{\Gamma}^2+\Delta\Omega^2)S_{Y_0Y_0}+S_{\bar{F}_{\rm th}\bar{F}_{\rm th}}
\end{equation}
in the limit of $|\Omega_1-\Omega_2|, \Gamma_1,\Gamma_2 \ll \Omega_1,\Omega_2$ where $S_{ Y_0Y_0}=$ V/2 with $V=1$ for coherent states and $V \sim 0.63$ for the entangled state with a measured 2 dB squeezing and $\bar{\Gamma}=\sqrt{(\Gamma_1^2+\Gamma_2^2)/2}$. $\Delta\Omega=\Omega_1-\Omega_2$  is the resonant frequency difference. The minimum force noise scales as $1/\alpha_c$ until thermal noise becomes comparable to the imprecision noise. We define the peak sensitivity as $\mathcal{S}=1/S_{\bar{F}_{\rm min}\bar{F}_{\rm min}}$. The entangled probes offer an improvement in the sensitivity by lowering the imprecision noise floor from the SNL by 2 dB.

In Fig.~\ref{fig:entangled_vs_classical} (d), we plot the joint homodyne PSDs of two sensors with a small resonant frequency difference (260 Hz) interrogated by entangled (red) or classical (black) probes each with 50 $\mu$W power. Fig.~\ref{fig:entangled_vs_classical} (e) shows that the dominant thermal noise $\sqrt{S_{\bar{F}_{\rm th}\bar{F}_{\rm th}}}=\sqrt{(S_{F_{\rm th}^{(1)}F_{\rm th}^{(1)}}+S_{F_{\rm th}^{(2)}F_{\rm th}^{(2)}})}/2$ around the resonant frequencies limits the peak force sensitivity for the joint force measurements with the entangled and classical probes, where $\sqrt{S_{F_{\rm th}^{(i)}F_{\rm th}^{(i)}}}=\sqrt{2\Gamma_i m_{\rm eff} k_B T}\sim10^{-15}$ N$/\sqrt{\text{Hz}}$ is thermal noise at each sensor. Figure \ref{fig:entangled_vs_classical}~(f) shows the minimum force noise (filled dots) at different probe power levels. The sensitivities for the entangled and classical probes both converge to the thermal noise limit as the probe power increases. However, the entangled probes can improve the sensing bandwidth as we next elaborate.

\subsection{Entanglement-enhanced measurement bandwidth}

The response of a mechanical oscillator to external stimuli is enhanced by its $Q$-factor, which boosts the transduction efficiency around the resonant frequency. Due to coupling to the thermal bath, the force sensitivity of single sensor is limited by the thermal noise which scales inversely as the Q/mass ratio. Recent development of ultra-high $Q$ mechanical resonators has enable dramatic improvements in force sensitivity at the cost of narrow sensitivity bandwidth~\cite{ghadimi2018elastic,beccari2022strained,maccabe2020nano,tsaturyan2017ultracoherent,hoj2021ultra}. By contrast, joint measurements undertaken by $M$ identical mechanical sensors with homogeneous resonant frequencies can improve force sensitivity by the SQL scaling of $1/\sqrt{M}$ while maintaining a bandwidth similar to that of a single sensor. Entangled probes can moreover increase the sensing bandwidth of sensor arrays, in analogy to recent demonstrations of squeezed-light-enhanced bandwidths for a microwave cavity sensor~\cite{malnou2019squeezed,backes2021quantum} and an optomechanical magnetometer~\cite{li2018quantum}. 

As shown in Fig.~\ref{fig:entangled_vs_classical} (e), it is evident that the bandwidth for entangled probes (red) is broadened compared to that for the classical probes (black). To quantify the bandwidth improvement by entangled probes, we define the 3-dB sensing bandwidth as $\mathcal{B}_{3\rm dB}\equiv\omega_{3{\rm dB}+}-\omega_{{3\rm dB}-}$, the width of the frequency band over which the force noise power is within a factor of 2 of the minimum, i.e., $ S_{\bar{F}\bar{F}}(\omega_{\rm 3dB\pm})=2S_{\bar{F}_{\rm min}\bar{F}_{\rm min}}$. Figure \ref{fig:entangled_vs_classical}~(f) shows the 3-dB sensing bandwidths (open dots) at different probe power levels and dashed lines correspond to theoretical predictions. The bandwidth approximately scales as $\alpha_c$ in the thermal noise-dominated regime and the entangled probes maintain a $20\%$ sensing bandwidth improvement. At large resonant frequency differences, for example in Fig.~\ref{fig:entangled_vs_classical} (c), the bandwidth is predominantly determined by the resonant frequency difference (1422 Hz) and increases marginally with the probe power. The bandwidth for the entangled probes is worse than that of the classical probes because the entangled state around the resonant frequencies is not optimized to account for the large disparity in mechanical transduction efficiencies of the two sensors. Frequency-dependent entangled states are required to fully exploit the advantage of quantum correlations.

\subsection{\label{subsec:tradeoff}Sensitivity-bandwidth product}

\begin{figure*}[th]
    \centering
    \includegraphics[width=1\textwidth]{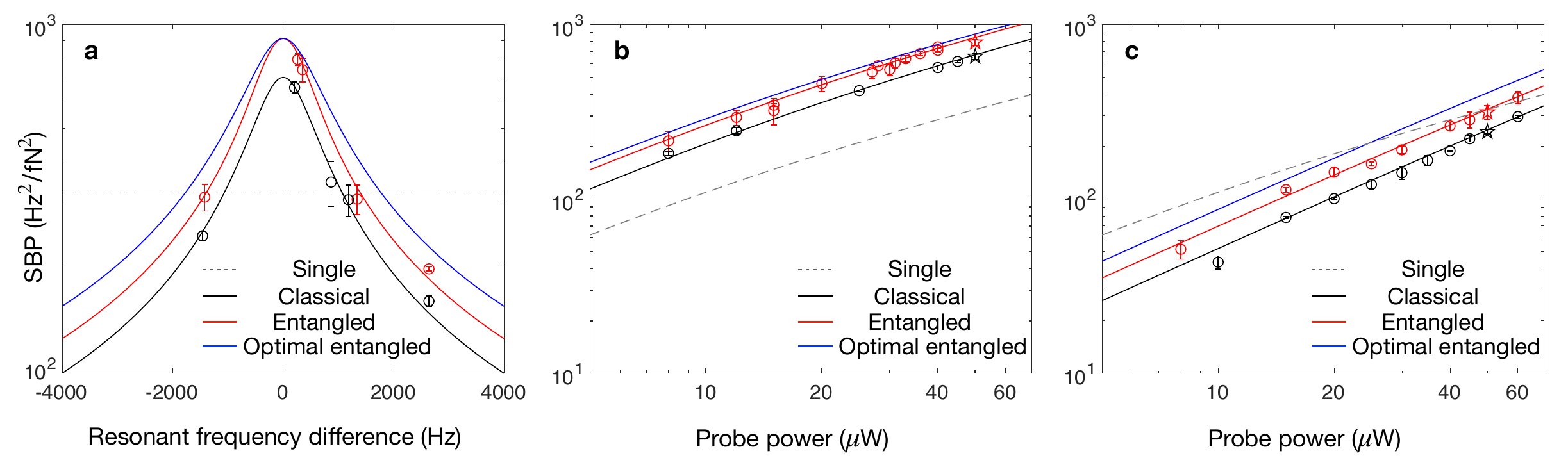}
    \caption{Sensitivity-bandwidth product as a figure of merit for joint force measurements. (a) SBP vs frequency differences suggesting a sensitivity-bandwidth tradeoff. Probe power is 50 $\mu$W for all measurements. (b, c) SBP vs probe power at resonant frequency difference of (b) 262 Hz and (c) 1422 Hz. In all figures, entangled probes: red; classical probes: black, optimal entangled probes: blue; dots: experimental data; solid lines: theory; dashed lines: single classical sensor. Error bars account for frequency difference drifting, theory fitting, and fluctuations in the noise power measurements.}
    \label{fig:SBP}
\end{figure*}

The previous two sets of data illustrate that sensors with a large resonant frequency difference enjoy a larger measurement bandwidth, and their sensitivity minimum can be enhanced by entangled probes. Conversely, sensors with a small resonant frequency difference present higher sensitivity while entangled probes can enlarge the measurement bandwidth. To highlight this feature, we first display in Fig.~\ref{fig:frequency_difference} (a) the homodyne PSDs acquired by entangled probes with 50 $\mu$W power at four resonant frequency differences. Figure~\ref{fig:frequency_difference} (b) then visualizes the dependence of the force noise and bandwidth on the resonant frequency difference. Figure~\ref{fig:frequency_difference} (c) and (d) depict the minimum force noise and 3-dB bandwidths associated with the entangled (red lines) and classical (black lines) probes at different frequency differences, showing a good agreement between theory and experiment. The bandwidth with entangled light approaches the performance of optimal entangled state (blue line) near zero resonant frequency difference but drops below the bandwidth of both optimal entangled and classical light at large resonant frequency differences while the minimum force noise using entangled light coincides with the minimum force noise achieved by the optimal entangled state. As a comparison, we also show the theoretical minimum force noise and 3-dB bandwidth of a single sensor (dashed gray lines) probed with classical light. The minimum force noise for two sensors with similar resonant frequencies is reduced by about $1/\sqrt{2}$ as compared to that of a single sensor. However, the peak sensitivity of two sensors with a large resonant frequency difference is worse than that of a single sensor as joint imprecision noise is dominant over thermal noise. The bandwidth of two sensors, however, is always larger than that of a single sensor and increases with the resonant frequency difference.

At this juncture, we introduce the sensitivity-bandwidth product (SBP)~\cite{korobko2017beating} to assess the overall performance of the joint force measurement. The SBPs of the classical and optimal entangled state are given by (see Supplemental Material):
\begin{equation}\label{eq:SBP}
   \mathcal{S}\times\mathcal{B}_{\rm 3dB} \approx\frac{\beta\alpha_c}{\omega_{\rm min}m_{\rm eff}\sqrt{S_{\rm Y_0Y_0}}}\sqrt{\frac{1}{S_{\bar{F}_{\rm min}\bar{F}_{\rm min}}}}.
\end{equation}
The SBP of the entangled probe in our experiment is suboptimal and lies in between the ones for the classical and optimal entangled probes.  Figure~\ref{fig:SBP} (a) shows the SBP at various resonant frequency differences for the classical (black), entangled (red) and optimal entangled probes (blue). The probe power is fixed at around 50 $\mu$W at each sensor for both the classical and entangled probes. SPBs decrease with respect to the resonant frequency difference as expected from Eq.~(\ref{eq:SBP}) that joint imprecision noise increases with resonant frequency differences. We also plot the SBP for a classical single sensor (dashed line) probed with 50 $\mu$W as a comparison. Beyond certain resonant frequency differences, the SBP for entangled probes can even drop below that of a single sensor. In the small resonant frequency difference scenario, the SBP of the entangled probes in our experiment is on par with that of the optimal entangled probes, surpassing the SBP of classical probes by a factor of $1/\sqrt{V}\sim 1.25$. We plot the SBPs against different probe power levels at two resonant frequency differences of 262 Hz and 1422 Hz in Fig.~\ref{fig:SBP} (b,c). The SBPs of two sensors increase with respect to the square root of the mean photon number $\alpha_c$ in the thermal noise dominant regime as shown in Fig.~\ref{fig:SBP} (b) and the mean photon number $\alpha_c^2$ in the imprecision noise dominant regime as shown in Fig.~\ref{fig:SBP} (c).

\begin{figure*}[thb!]
    \centering
    \includegraphics[width=1\textwidth]{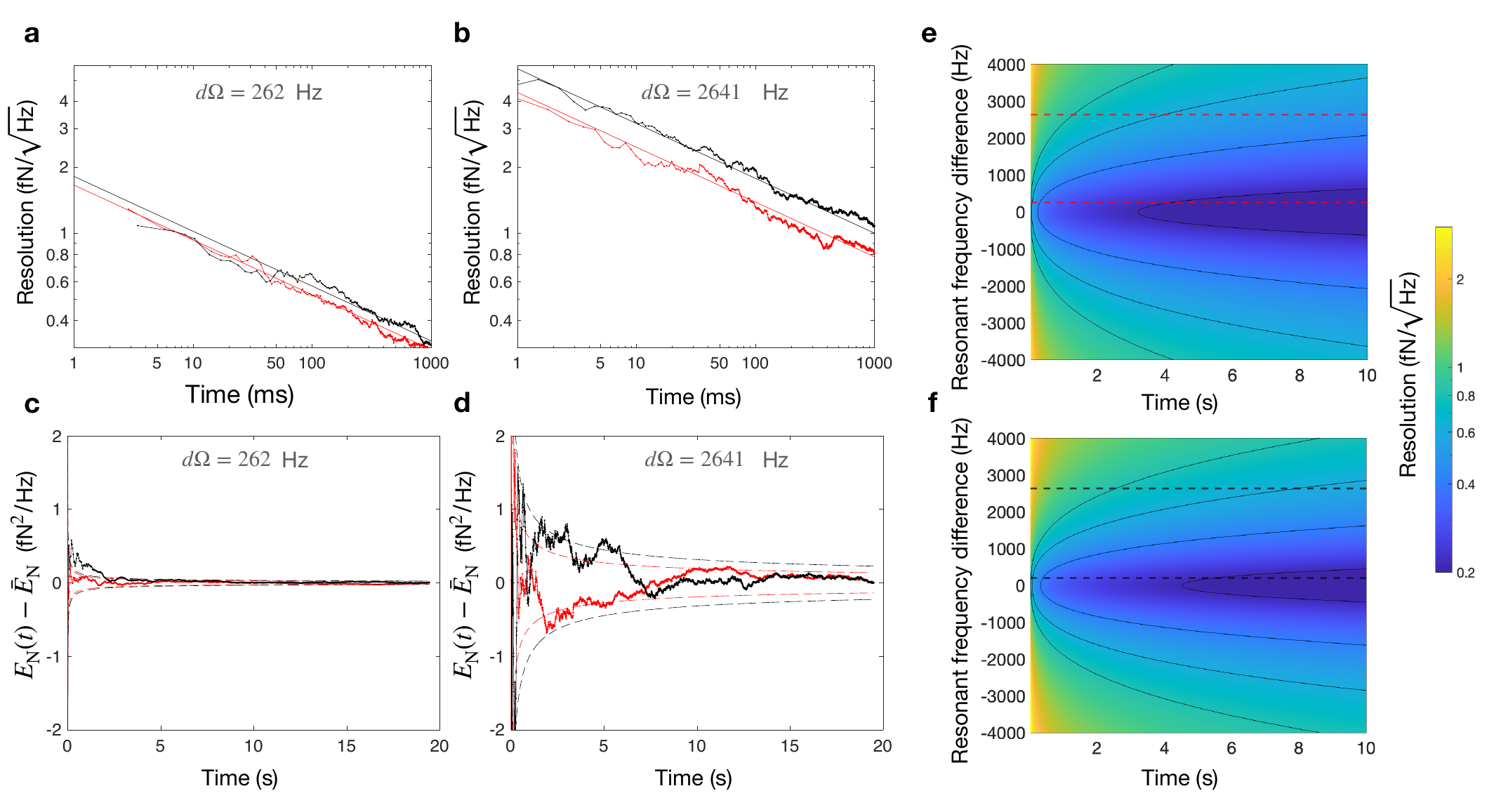}
    \caption{Incoherent force sensing. (a, b) Force resolution vs integration time based on $50~\mu$W of entangled (red) and classical (black) probes on sensors with a resonant frequency difference of (a) 262 Hz and (b) 2641 Hz. (c, d) Estimation of noise force power at different time for $50~\mu$W entangled (red) and classical (black) probes with resonant frequency difference (c) 262 Hz and (d) 2641 Hz. Dashed lines are the force resolution given by (a, b). (e, f) Simulated force resolution with different resonant frequency difference at different time for (e) entangled and (f) classical probe. Theory lines in (a, b) are marked as the dashed lines in (e, f). Black lines are the contours of resolution at $\{0.1,0.21,0.41,0.61,0.81\}$ $\text{fN}/\sqrt{\text{Hz}}$.}
    \label{fig:incoherent}
\end{figure*}
\section{\label{sec:incoherent}Entanglement-enhanced incoherent force sensing}

Optomechanical sensors have been exploited in detecting weak incoherent forces embedded in a thermal-noise background~\cite{gavartin2012hybrid,harris2013minimum}, a regime pertinent to dark matter searches~\cite{carney2021ultralight,brady2022DQSoptomech}. Joint measurements taken by multiple sensors can increase the SNR by lowering the measurement noise, thereby enhancing the resolution in incoherent force sensing. Entangled probes, in this regard, can further improve incoherent force sensing by increasing the measurement bandwidth (resolution) in the thermal-noise-dominant (imprecision-noise-dominant) scenario. Following Ref.~\cite{gavartin2012hybrid,harris2013minimum}, we choose the energy estimator as a performance metric and use it to demonstrate a quantum advantage in measurement of uncorrelated, incoherent forces.

 We define the equivalent force spectral resolution (EFSR) as $\delta F_{\rm N}=\sqrt{\delta E_{\rm N}(t)}$, i.e., the square root of the standard deviation of the overall force noise within its 3-dB bandwidth, which includes the thermal force and an equivalent force noise contributed by the imprecision noise of the probes, where $E_{\rm N}(t)=\int^t_0 dt F_{\rm N}(t)^2/t$ is the noise force energy averaged over $t$ seconds. An incoherent force is detectable only if the EFSP is finer than its standard deviation. Figure~\ref{fig:incoherent} (a) shows the EFSR versus integration time for the entangled (red) and classical (black) probes each carrying 50 $\mu$W of power. The resonant frequency difference is 262 Hz so the measurement is dominated by the thermal force, resulting in similar force resolutions for the entangled and classical probes. The data corroborate the force-resolution scaling of $t^{-1/4}$ for both types of probes, as predicted by theory\cite{gavartin2012hybrid}. However, the entangled probes offer a larger measurement bandwidth (enabling accelerate search for unknown signals, in the same spirit of squeezed-light-enhanced dark matter search based on microwave cavity sensors~\cite{malnou2019squeezed,backes2021quantum}). Figure~\ref{fig:incoherent} (b) shows the time dependence of the force resolution for sensors with 2641-Hz resonant frequency difference. The entangled probes reduce the integration time by $60\%$ over that of the classical probes in arriving at the same force resolution. Figure~\ref{fig:incoherent} (e, f) present the simulation result for the force resolution vs the resonant frequency difference and integration time attained by entangled (e) and classical probes (f). The dashed lines in Fig.~\ref{fig:incoherent} (e, f) correspond to the theory curves in Fig.~\ref{fig:incoherent} (a, b). The estimated force power ($E_{\rm N}(t)-\bar{E}_{\rm N}$) are shown in Fig.~\ref{fig:incoherent} (c, d) and converge to zero at long integration time, where $\bar{E}_{\rm N}$ is the mean noise force power. The dashed lines correspond to the force resolution in Fig.~\ref{fig:incoherent}(a,b). A stationary incoherent force signal can only be resolved when its energy within the detection bandwidth exceeds the noise power uncertainty at a given averaging time. Without loss of generality, we only plot the estimation for the total noise force. An example for the signal-force estimation is presented in the Supplementary Material.

\section{\label{sec:Discussions}Conclusions}

In conclusion, we have experimentally demonstrated entanglement-enhanced joint force measurements with two optomechanical sensors. Sensitivity and bandwidth enhancement enabled by entangled probes are predicted and observed. Specifically, optomechanical sensors jointly probed by entangled light generated from a passive beam splitter array with squeezed-light input outperform the same sensors probed by classical laser light in force resolution and measurement bandwidth. A full performance analysis of more sensors including backaction effect and resource counting is given in the accompanying theoretical paper~\cite{brady2022DQSoptomech}.
Our work opens a new avenue for ultraprecise measurements with an array of quantum-enhanced sensors, for applications ranging from inertial navigation to acoustic imaging, to searches for new physics.

\section*{Acknowledgments}
This work is supported by the Office of Naval Research Grant No. N00014-19-1-2190, National Science Foundation Grants No. 2040575, 2134830, and U.S. Department of Energy, Office of Science, National Quantum Information Science Research Centers, Superconducting Quantum Materials and Systems Center (SQMS) under the contract No. DE-AC02-07CH11359. Q.Z. also acknowledges support from Defense Advanced Research Projects Agency (DARPA) under Young Faculty Award (YFA) Grant No. N660012014029 and NSF CAREER Award CCF-2142882.
\appendix

%\bibliography{myref}

%apsrev4-2.bst 2019-01-14 (MD) hand-edited version of apsrev4-1.bst
%Control: key (0)
%Control: author (8) initials jnrlst
%Control: editor formatted (1) identically to author
%Control: production of article title (0) allowed
%Control: page (0) single
%Control: year (1) truncated
%Control: production of eprint (0) enabled
%



\section*{Supplementary material}

# Supplemental Material for Entanglement-Enhanced Optomechanical Sensing

Yi Xia,[1,2,*] Aman R. Agrawal,[1] Christian M. Pluchar,[1] Anthony J. Brady,[3]
Zhen Liu,[4] Quntao Zhuang,[3,1,5] Dalziel J. Wilson,[1] and Zheshen Zhang[2,3,1,6,†]

[1]*James C. Wyant College of Optical Sciences, The University of Arizona, Tucson, Arizona 85721, USA*
[2]*Department of Materials Science and Engineering,*
*The University of Arizona, Tucson, Arizona 85721, USA*
[3]*Department of Electrical and Computer Engineering,*
*The University of Arizona, Tucson, Arizona 85721, USA*
[4]*School of Physics and Astronomy, University of Minnesota, Minneapolis, MN 55455, USA*
[5]*Department of Electrical and Computer Engineering,*
*University of Southern California, Los Angeles, CA 90089, USA*
[6]*Department of Electrical Engineering and Computer Science,*
*University of Michigan, Ann Arbor, MI 48109, USA*

## I. THEORETICAL FRAMEWORK

### A. Notation convention

Following the standard definition, the Fourier transform of an operator (ignoring the hat for all operators) is:

$$X(\omega) = \int_{-\infty}^{+\infty} X(t) e^{i\omega t} dt. \tag{S1}$$

The Hermitian conjugate of an operator in frequency domain is defined as $X^\dagger(\omega) = [\int_{-\infty}^{+\infty} X(t) e^{i\omega t} dt]^\dagger$. For Hermitian operator $X^\dagger(t) = X(t)$, we immediately find $X^\dagger(\omega) = X(-\omega)$.

The quantum power spectral density (PSD) with stationary statistics is defined as [1]:

$$\begin{aligned} S_{XX}(\omega) &= \int_{-\infty}^{+\infty} \langle X(t) X(0) \rangle e^{i\omega t} dt \\ &= \int_{-\infty}^{+\infty} \langle X(\omega) X(\omega') \rangle d\omega' \end{aligned} \tag{S2}$$

We employ the symmetrized PSD $\bar{S}_{XX}(\omega) = (S_{XX}(\omega) + S_{XX}(-\omega))/2$ in the following context which relates to the homodyne detection of optical phase quadrature and keep the notation $S_{XX}(\omega)$ for simplicity. The noise PSD of the phase quadrature $Y$ of phase squeezed light (assuming flat squeezing spectrum) is given by:

$$S_{YY}(\omega) = \frac{V}{2}, \tag{S3}$$

where $V = e^{-2r}$ for pure squeezed states. $r = 0$ corresponds to the vacuum noise.

The noise PSD of the thermal force $F_{\text{th}}$ (high temperature classical limit) driving the mechanical membrane is given by:

$$S_{F_{\text{th}} F_{\text{th}}}(\omega) = 2\Gamma m_{\text{eff}} k_B T, \tag{S4}$$

---

[*] yixia.light@gmail.com
[†] zszh@umich.edu

where $\Gamma$ is mechanical damping rate, $m_{\text{eff}}$ is effective mass of mechanical mode. $k_B$ is Boltzmann's constant and $T$ is the environmental temperature.

## B. Optical readout of mechanical displacement

Considering an optomechanical system where the resonance frequency of intra cavity field $a(t) = \bar{a} + \delta a$ is dispersively coupled to the displacement of mechanical resonator $x(t)$, the linearized equations of motion are given by [2]:

$$\delta \dot{a} = \left(i\Delta - \frac{\kappa}{2}\right)\delta a + iG\bar{a}\delta x + \sqrt{\kappa}\delta a_{\text{in}}, \tag{S5}$$

$$m_{\text{eff}}(\delta\ddot{x} + \Gamma\delta\dot{x} + \Omega^2\delta x) = F_{\text{th}} + F_{\text{sig}} + \hbar G\bar{a}(\delta a + \delta a^\dagger). \tag{S6}$$

Here, $G$ is the optomechanical coupling strength. $m_{\text{eff}}, \Gamma, \Omega$ are the effective mass, damping rate, and resonance frequency of the interested mechanical mode respectively. $\kappa$ is the optical cavity coupling rate to the external field $a_{in}$. The steady state intracavity photon number is:

$$\bar{a}^2 = \frac{\kappa}{\Delta^2 + (\kappa/2)^2}\bar{a}_{\text{in}}^2 \tag{S7}$$

In the experimentally relevant situation of resonant probing ($\Delta \approx 0$), bad cavity limit ($\kappa \gg \Omega$) and negligible backaction noise (last term in Eq. (S6)), we can solve the coupled equations in the frequency domain, yielding

$$\delta a(\omega) = \frac{2}{\sqrt{\kappa}}\delta a_{\text{in}}(\omega) + iG\frac{4}{\kappa\sqrt{\kappa}}\bar{a}_{\text{in}}\delta x(\omega) \tag{S8}$$

$$\delta x(\omega) = \chi(\omega)(F_{\text{th}} + F_{\text{sig}}), \tag{S9}$$

where

$$\chi(\omega) = \frac{1/m_{\text{eff}}}{\Omega^2 - \omega^2 + i\Gamma\omega} \tag{S10}$$

is the mechanical susceptibility. Using the input-output relation $\delta a_{\text{out}}(\omega) + \delta a_{\text{in}}(\omega) = \sqrt{\kappa}\delta a(\omega)$, the quantum fluctuation of output field is given by:

$$\delta a_{\text{out}}(\omega) = \delta a_{\text{in}}(\omega) + i\frac{4G}{\kappa}\bar{a}_{\text{in}}\delta x(\omega). \tag{S11}$$

It is thus evident that the displacement information, $\delta x$, is encoded into the phase of output field. We define the phase quadrature of the optical field as $Y_{\text{out}}(t) = (\delta a_{\text{out}}(t) - \delta a_{\text{out}}^\dagger(t))/i\sqrt{2}$. We use homodyne detection to measure the phase quadrature of output probe:

$$Y_{\text{out}}(\omega) = Y_{\text{in}}(\omega) + \alpha\beta\chi(\omega)(F_{\text{th}} + F_{\text{sig}}), \tag{S12}$$

where we define the optomechanical coupling efficiency $\beta \equiv 4\sqrt{2}G/\kappa$ and the amplitude of input probe light $\alpha \equiv \bar{a}_{\text{in}}$. $\alpha$ is assumed real for simplicity. Without signals, the measured PSD and its corresponding minimum detectable force PSD are calculated to be

$$S_{Y_{\text{out}}Y_{\text{out}}} = S_{Y_{\text{in}}Y_{\text{in}}} + \alpha^2\beta^2|\chi|^2 S_{F_{\text{th}}F_{\text{th}}}, \tag{S13}$$

$$S_{F_{\text{min}}F_{\text{min}}} = S_{Y_{\text{in}}Y_{\text{in}}}/(\alpha^2\beta^2|\chi|^2) + S_{F_{\text{th}}F_{\text{th}}}. \tag{S14}$$

It is possible to improve force sensitivity at off-resonance frequencies and the sensing bandwidth near the resonance frequency by engineering the quantum noise properties of the probe light, i.e., $S_{Y_{\text{in}}Y_{\text{in}}}(\omega)$, using, e.g., squeezed light to

surpass the shot noise limit as demonstrated in recent experiments.

### C. Two mode entangled optical probe

Let us now consider two independent optomechanical sensors. The phase quadratures of the two output probes are given by

$$Y_{\text{out}}^{(i)}(\omega) = Y_{\text{in}}^{(i)}(\omega) + \alpha_i \beta_i \chi_i(\omega) \left( F_{\text{th}}^{(i)} + F_{\text{sig}}^{(i)} \right), \tag{S15}$$

where $i \in \{1, 2\}$ denotes the sensor index. To estimate global properties across the two sensors, e.g., the average of the two force signals $\bar{F}_{\text{sig}} = \left( F_{\text{sig}}^{(1)} + F_{\text{sig}}^{(2)} \right)/2$, we can choose the force estimator as

$$\begin{aligned}
F_{\text{est}} &= \frac{1}{2} \left( \frac{Y_{\text{out}}^{(1)}}{\alpha_1 \beta_1 \chi_1} + \frac{Y_{\text{out}}^{(2)}}{\alpha_2 \beta_2 \chi_2} \right) \\
&= \frac{\sqrt{(\alpha_1 \beta_1 |\chi_1|)^2 + (\alpha_2 \beta_2 |\chi_2|)^2}}{2 \alpha_1 \beta_1 \chi_1 \alpha_2 \beta_2 \chi_2} Y_0 + \bar{F}_{\text{th}} + \bar{F}_{\text{sig}}.
\end{aligned} \tag{S16}$$

Here, we define the averaged thermal force $\bar{F}_{\text{th}} = \left( F_{\text{th}}^{(1)} + F_{\text{th}}^{(2)} \right)/2$, and a normalized quadrature operator $Y_0 = \eta_1 Y_{\text{in}}^{(1)} + \eta_2 Y_{\text{in}}^{(2)}$, where the coefficients are

$$\begin{aligned}
\eta_1(\omega) &= \frac{\alpha_2 \beta_2 \chi_2}{\sqrt{(\alpha_1 \beta_1 |\chi_1|)^2 + (\alpha_2 \beta_2 |\chi_2|)^2}} \\
\eta_2(\omega) &= \frac{\alpha_1 \beta_1 \chi_1}{\sqrt{(\alpha_1 \beta_1 |\chi_1|)^2 + (\alpha_2 \beta_2 |\chi_2|)^2}}.
\end{aligned} \tag{S17}$$

The joint force-noise PSD is determined by the averaged thermal noise $S_{\bar{F}_{\text{th}} \bar{F}_{\text{th}}} = \left( S_{F_{\text{th}}^{(1)} F_{\text{th}}^{(1)}} + S_{F_{\text{th}}^{(2)} F_{\text{th}}^{(2)}} \right)/4$ and joint quadrature noise

$$S_{Y_0 Y_0} = |\eta_1|^2 S_{Y_{\text{in}}^{(1)} Y_{\text{in}}^{(1)}} + \eta_1 \eta_2^* S_{Y_{\text{in}}^{(1)} Y_{\text{in}}^{(2)}} + \eta_1^* \eta_2 S_{Y_{\text{in}}^{(2)} Y_{\text{in}}^{(1)}} + |\eta_2|^2 S_{Y_{\text{in}}^{(2)} Y_{\text{in}}^{(2)}}. \tag{S18}$$

We now show how to engineer the quantum correlation between the two probes such that $S_{Y_0 Y_0} < 1$ to beat the shot noise limit. The entangled probe in our experiment is generated by splitting a squeezed light into two arms, similar to the entangled light used in the recent experimental demonstration of distributed quantum sensing. The squeezed quadrature $b$ mixes with quadrature $c$ of the vacuum port of the beam splitter with transmissivity $v_1$ and reflectivity $v_2$, where $|v_1|^2 + |v_2|^2 = 1$. The noise PSDs of the two quadratures are $S_{bb} = e^{-2r}/2$ and $S_{cc} = 1/2$ respectively. The output beams from the beam splitter are entangled and are used to probe the optomechanical sensors. The probes' phase quadratures are expressed as:

$$Y_{\text{in}}^{(1)} = v_1 b - v_2 c, \tag{S19}$$
$$Y_{\text{in}}^{(2)} = v_2 b + v_1 c. \tag{S20}$$

The associated PSDs and cross spectral densities can be derived as follows:

$$\begin{aligned}
S_{Y_{\text{in}}^{(1)} Y_{\text{in}}^{(1)}} &= v_1^2 e^{-2r}/2 + v_2^2/2 \\
S_{Y_{\text{in}}^{(2)} Y_{\text{in}}^{(2)}} &= v_2^2 e^{-2r}/2 + v_1^2/2 \\
S_{Y_{\text{in}}^{(1)} Y_{\text{in}}^{(2)}} &= v_1 v_2 e^{-2r}/2 - v_1 v_2/2
\end{aligned} \tag{S21}$$

Plugging Eq. (S21) into Eq. (S18) yields the join quadrature noise PSD.

Before discussing the entangled probes, we first present the case with classical probes to set the SQL for quantum metrology. Using laser light for the probes, one has $S_{bb} = 1/2$, and the joint noise PSD

$$S_{\bar{F}\bar{F}}^c = \frac{(\alpha_1\beta_1|\chi_1|)^2 + (\alpha_2\beta_2|\chi_2|)^2}{(2\alpha_1\beta_1|\chi_1|\alpha_2\beta_2|\chi_2|)^2} \frac{1}{2} + S_{\bar{F}_{\text{th}}\bar{F}_{\text{th}}} \tag{S22}$$

In a simple case of $\alpha_1 = \alpha_2 = \alpha, \beta_1 = \beta_2 = \beta, \chi_1 = \chi_2 = \chi$, the joint force-noise floor is

$$\frac{1}{2}\left[\left|\frac{1}{\alpha\beta|\chi|}\right|^2 + S_{F_{\text{th}}F_{\text{th}}}\right],$$

3 dB below the noise floor for a single sensor. The force sensitivity is thus improved by $1/\sqrt{2}$ as guaranteed by the central limit theorem.

For the entangled probes, the quantum correlations are generally frequency dependent as shown in Eq. (S17). However, when $\chi_1(\omega) = \chi_2(\omega)$, Eq. (S17) become frequency independent:

$$\eta_1 = \frac{\alpha_2\beta_2}{\sqrt{(\alpha_1\beta_1)^2 + (\alpha_2\beta_2)^2}}$$

$$\eta_2 = \frac{\alpha_1\beta_1}{\sqrt{(\alpha_1\beta_1)^2 + (\alpha_2\beta_2)^2}}. \tag{S23}$$

We can engineer the entanglement by setting $v_1 = \eta_1$ and $v_2 = \eta_2$ such that the effective quadrature operator $Y_0 = b$. The joint force-noise PSD then simply reads

$$S_{\bar{F}\bar{F}}^o = \frac{(\alpha_1\beta_1)^2 + (\alpha_2\beta_2)^2}{(2|\chi|\alpha_1\beta_1\alpha_2\beta_2)^2} \frac{e^{-2r}}{2} + S_{\bar{F}_{\text{th}}\bar{F}_{\text{th}}}, \tag{S24}$$

showing that the reduced quantum noise due to the entangled probes is equivalent to that of the single mode squeezed vacuum state at the input of the beam splitter.

When $\chi_1 \neq \chi_2$, the optimal entangled state for the probes is in general frequency dependent as the mechanical transduction efficiencies $\propto \chi_i(\omega)$ for the two sensors are disparate across their sideband frequencies. To produce the optimal entangled states for the probes, frequency-dependent beam splitting ratios, e.g., $v_1(\omega) = |\eta_1(\omega)|$ and $v_2(\omega) = |\eta_2(\omega)|$, would be needed. In this case, the joint quadrature noise PSD becomes

$$S_{Y_0Y_0} = \left[\frac{e^{-2r}}{2} + v_1^2 v_2^2(1 - \cos(\theta_1 - \theta_2))(1 - e^{-2r})\right], \tag{S25}$$

where we define the transduction phase as $\theta_i(\omega) = \text{Arg}(\chi_i(\omega))$. The transduction phase difference $\theta_1 - \theta_2$, however, degrades the performance as the second term in Eq. (S25) increases the total noise power. The transduction phase difference can be compensated by delaying either one of the force signal $F_{\text{sig}}^{(i)} \to F_{\text{sig}}^{(i)} e^{i\phi}$ or probe phase quadrature $Y_{\text{in}}^{(i)} \to Y_{\text{in}}^{(i)} e^{i\phi}$, making $\cos(\theta_1 - \theta_2 - \phi) = 1$. In doing so, the joint force-noise PSD is found to be

$$S_{\bar{F}\bar{F}}^o = \frac{(\alpha_1\beta_1|\chi_1|)^2 + (\alpha_2\beta_2|\chi_2|)^2}{(2\alpha_1\beta_1|\chi_1|\alpha_2\beta_2|\chi_2|)^2} \frac{e^{-2r}}{2} + S_{\bar{F}_{\text{th}}\bar{F}_{\text{th}}} \tag{S26}$$

In comparison with the force-noise PSD for the classical probes in Eq. (S22), one finds that Eq. (S26) indeed quantifies the performance achieved by the optimal entangled probes.

### D. Practical entangled probes in experiment

Due to technical constraints, the splitting ratio in the experiment is frequency independent. Our objective is to generate a frequency-independent entangled state that minimizes the measurement noise in the shot-noise-dominant

region. At far-off-resonance frequencies $|\omega - \Omega_i| \gg \Gamma_i$, the shot noise dominates the measurement and the mechanical response functions converge $\chi_1 \approx \chi_2$. We could choose the same entangled state for the probes as the one for identical mechanical susceptibilities that led to the minimized force-noise PSD described by Eq. (S24), with the splitting ratio

$$v_1 = \frac{\alpha_2 \beta_2}{\sqrt{(\alpha_1 \beta_1)^2 + (\alpha_2 \beta_2)^2}}$$
$$v_2 = \frac{\alpha_1 \beta_1}{\sqrt{(\alpha_1 \beta_1)^2 + (\alpha_2 \beta_2)^2}}. \tag{S27}$$

The joint force-noise PSD is:

$$S^{\mathrm{e}}_{\bar{F}\bar{F}} = \frac{\alpha_1^2 \beta_1^2 + \alpha_2^2 \beta_2^2}{(2\alpha_1 \beta_1 \alpha_2 \beta_2)^2} \left[ \frac{v_1^2 S_{Y^{(1)}_{\mathrm{in}} Y^{(1)}_{\mathrm{in}}}}{|\chi_1|^2} + \frac{v_2^2 S_{Y^{(2)}_{\mathrm{in}} Y^{(2)}_{\mathrm{in}}}}{|\chi_2|^2} + \frac{2 v_1 v_2}{|\chi_1||\chi_2|} S_{Y^{(1)}_{\mathrm{in}} Y^{(2)}_{\mathrm{in}}} \cos(\theta_1 - \theta_2 - \phi) \right] + S_{\bar{F}_{\mathrm{th}} \bar{F}_{\mathrm{th}}} \tag{S28}$$

We note that the power at the optical carrier frequency is also split by the beam splitter in generating the entangled probes. Specifically, the squeezed light in our experiment comprises a coherent state portion with amplitude $\alpha_c$ at carrier frequency and squeezed vacuum states at sideband frequencies, rendering $\alpha_1 = v_1 \alpha_c$, $\alpha_2 = v_2 \alpha_c$. Substituting $\alpha_1$ and $\alpha_2$ into Eq. (S27), $v_1, v_2$ can be derived in terms of the transduction efficiencies $\beta_1, \beta_2$ at two sensors:

$$v_1 = \sqrt{\frac{\beta_2}{\beta_1 + \beta_2}} \tag{S29}$$

$$v_2 = \sqrt{\frac{\beta_1}{\beta_1 + \beta_2}}. \tag{S30}$$

The above equations indicate that a lower optomechanical coupling efficiency requires more probe power to balance the overall force detection efficiency at two sensors. The joint force PSD for the experimental entangled probes can then be derived as

$$S^{\mathrm{e}}_{\bar{F}\bar{F}} = \frac{(\beta_1 + \beta_2)^2}{(2\beta_1 \beta_2)^2 \alpha_c^2} \left[ \frac{v_1^2 S_{Y^{(1)}_{\mathrm{in}} Y^{(1)}_{\mathrm{in}}}}{|\chi_1|^2} + \frac{v_2^2 S_{Y^{(2)}_{\mathrm{in}} Y^{(2)}_{\mathrm{in}}}}{|\chi_2|^2} + \frac{2 v_1 v_2}{|\chi_1||\chi_2|} S_{Y^{(1)}_{\mathrm{in}} Y^{(2)}_{\mathrm{in}}} \cos(\theta_1 - \theta_2 - \phi) \right] + S_{\bar{F}_{\mathrm{th}} \bar{F}_{\mathrm{th}}}. \tag{S31}$$

The joint force-noise PSDs of the classical light and optimal frequency dependent entangled light are shown below for comparison:

$$S^{\mathrm{c}}_{\bar{F}\bar{F}} = \frac{(\beta_1 + \beta_2)^2}{(2\beta_1 \beta_2)^2 \alpha_c^2} \left[ \frac{v_1^2}{|\chi_1|^2} + \frac{v_2^2}{|\chi_2|^2} \right] \frac{1}{2} + S_{\bar{F}_{\mathrm{th}} \bar{F}_{\mathrm{th}}},$$
$$S^{\mathrm{o}}_{\bar{F}\bar{F}} = \frac{(\beta_1 + \beta_2)^2}{(2\beta_1 \beta_2)^2 \alpha_c^2} \left[ \frac{v_1^2}{|\chi_1|^2} + \frac{v_2^2}{|\chi_2|^2} \right] \frac{e^{-2r}}{2} + S_{\bar{F}_{\mathrm{th}} \bar{F}_{\mathrm{th}}}. \tag{S32}$$

### E. Sensitivity-bandwidth product

In many sensing scenarios, the bandwidths of the signals to measure are typically much broader than the mechanical linewidth. For example, accelerometers usually operate below their fundamental resonance frequencies where the mechanical response displays an almost flat frequency response at the cost of a reduced force sensitivity from the resonance frequencies. To quantify the overall performance of optomechanical sensors that account for both the sensitivity and measurement bandwidth, we next introduce the sensitivity-bandwidth product (SPB) as a figure of merit [3].

We first derive the SPBs for classical and optimal entangled probes, building on their joint force-noise PSDs formulated in Eq. (S32). Without loss of generality, the transduction efficiencies are intentionally set to similar ($\beta_1 \approx \beta_2 = \beta$)

in our experiment such that the entangled probes can be generated by evenly splitting the squeezed light into two arms ($v_1 = v_2 = 1/\sqrt{2}$). We take the derivatives of Eq. (S32) to find the minimum of the force-noise PSD $S_{\bar{F}_{\min}\bar{F}_{\min}}$ and its corresponding frequency $\omega_{\min}$:

$$\omega_{\min} = \frac{1}{2}\sqrt{-\Gamma_1^2 - \Gamma_2^2 + 4\bar{\Omega}^2 + \Delta\Omega^2} \approx \bar{\Omega} \tag{S33}$$

$$S_{\bar{F}_{\min}\bar{F}_{\min}} = \frac{m_{\text{eff}}^2(\beta_1+\beta_2)^2}{(2\beta_1\beta_2)^2\alpha_c^2}\frac{(8\bar{\Omega}^2 - \Gamma_1^2 - \Gamma_2^2)(\Gamma_1^2+\Gamma_2^2) + 2\Delta\Omega^2(8\bar{\Omega}^2+\Gamma_1^2+\Gamma_2^2)}{16}S_{Y_0 Y_0} + S_{\bar{F}_{\text{th}}\bar{F}_{\text{th}}}$$
$$\approx \frac{m_{\text{eff}}^2}{\beta^2\alpha_c^2}\bar{\Omega}^2(\bar{\Gamma}^2 + \Delta\Omega^2)S_{Y_0 Y_0} + S_{\bar{F}_{\text{th}}\bar{F}_{\text{th}}}. \tag{S34}$$

Here, $\bar{\Omega} = (\Omega_1+\Omega_2)/2$, $\bar{\Gamma} = \sqrt{(\Gamma_1^2+\Gamma_2^2)/2}$ and $\Delta\Omega = \Omega_1 - \Omega_2$. $S_{Y_0 Y_0} = 1/2$ ($e^{-2r}/2$) is the quadrature-noise PSD for the classical (optimal entangled) probes. The first term in Eq. (S34) represents the joint imprecision noise while the second term describes the averaged thermal noise. The entangled probes reduce the imprecision noise by a factor of $e^{-2r}$ compared to the classical probes.

We next define the force sensitivity as

$$1/S_{\bar{F}\bar{F}} = \frac{16P}{16PS_{\bar{F}_{\text{th}}\bar{F}_{\text{th}}} + (8\bar{\Omega}^2-\Gamma_1^2-\Gamma_2^2)(\Gamma_1^2+\Gamma_2^2) + 2\Delta\Omega^2(8\bar{\Omega}^2+\Gamma_1^2+\Gamma_2^2) + 16(\Omega^2-\omega_{\min}^2)^2}, \tag{S35}$$

where $P = \beta^2\alpha_c^2/m_{\text{eff}}^2 S_{Y_0 Y_0}$. The peak sensitivity $\mathcal{S}$ is then given by

$$\mathcal{S} = \frac{16P}{16PS_{\bar{F}_{\text{th}}\bar{F}_{\text{th}}} + (8\bar{\Omega}^2-\Gamma_1^2-\Gamma_2^2)(\Gamma_1^2+\Gamma_2^2) + 2\Delta\Omega^2(8\bar{\Omega}^2+\Gamma_1^2+\Gamma_2^2)} = 1/S_{\bar{F}_{\min}\bar{F}_{\min}}. \tag{S36}$$

The 3-dB bandwidth, $\mathcal{B}_{3\text{dB}} \equiv \omega_{3\text{dB}+} - \omega_{3\text{dB}-}$, is defined as the frequency range over which the force noise is within a factor of 2 of the minimum force noise, i.e.,

$$S_{\bar{F}\bar{F}}(\omega_{3\text{dB}\pm}) = 2S_{\bar{F}_{\min}\bar{F}_{\min}}. \tag{S37}$$

One can show that the 3-dB frequencies are given by

$$\omega_{3\text{dB}\pm} = \sqrt{\omega_{\min}^2 \pm \frac{\beta\alpha_c}{m_{\text{eff}}}\sqrt{\frac{S_{\bar{F}_{\min}\bar{F}_{\min}}}{S_{Y_0 Y_0}}}}, \tag{S38}$$

and the 3-dB bandwidth is

$$\mathcal{B}_{3\text{dB}} = \sqrt{\omega_{\min}^2 + \frac{\beta\alpha_c}{m_{\text{eff}}}\sqrt{\frac{S_{\bar{F}_{\min}\bar{F}_{\min}}}{S_{Y_0 Y_0}}}} - \sqrt{\omega_{\min}^2 - \frac{\beta\alpha_c}{m_{\text{eff}}}\sqrt{\frac{S_{\bar{F}_{\min}\bar{F}_{\min}}}{S_{Y_0 Y_0}}}}$$
$$\approx \frac{\beta\alpha_c}{\omega_{\min}m_{\text{eff}}}\sqrt{\frac{S_{\bar{F}_{\min}\bar{F}_{\min}}}{S_{Y_0 Y_0}}}. \tag{S39}$$

The minimum force noise and 3-dB bandwidth for the entangled probes in the experiment are numerically calculated using Eq. (S31) and plotted together with those for the classical and optimal entangled probes in Fig. S1. The optimal entangled probes offer the largest bandwidth compared to both the experimental entangled and classical probes. The experimental entangled probes approach the bandwidth of the optimal entangled probes at the vicinity of zero frequency difference but offer less bandwidth as compared to the classical probes at large frequency differences. The force sensitivity is limited by the thermal noise near zero frequency difference for all types of probes. Nevertheless,

6

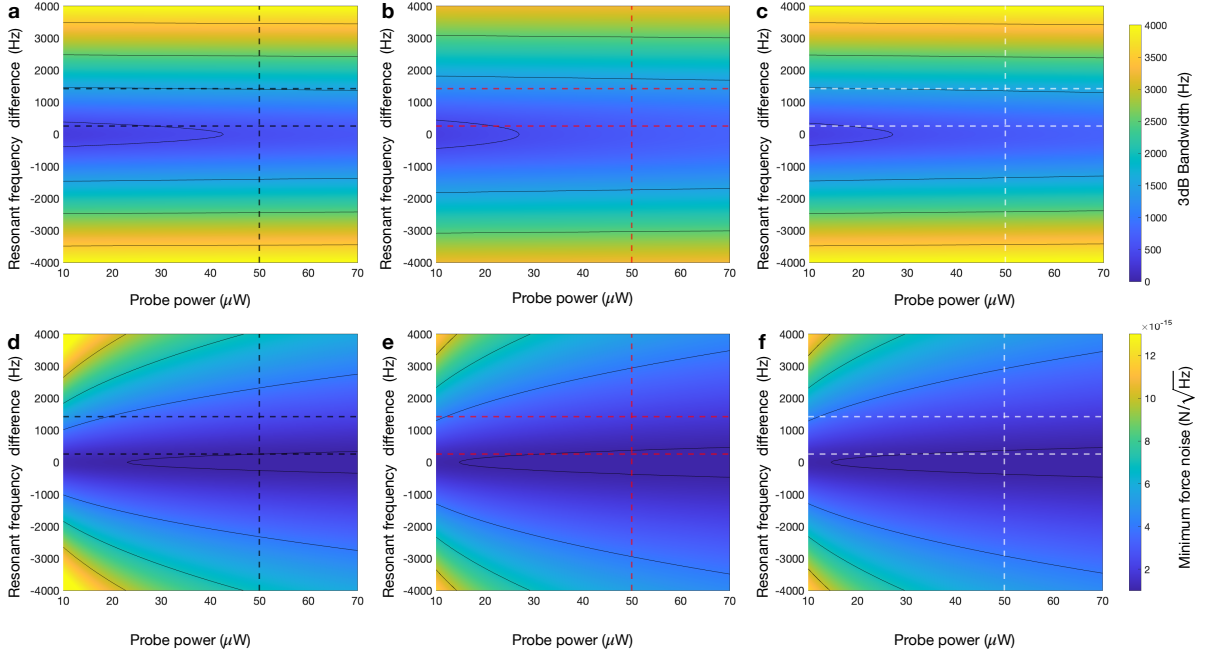

FIG. S1. Simulation of bandwidth and force sensitivity at different resonance frequency differences and probe power. (a-c) Bandwidth using classical, practical entangled, and optimal entangled probe. (d-f) Sensitivity using classical, practical entangled, and optimal entangled probe. Black solid lines: contours. Horizontal black (red) dashed lines: same as bandwidth (dashed lines) and sensitivity (solid lines) using classical (black) and practical entangled probe (red) in main text Fig. 2 (c,f). Vertical black (red) dashed lines: same as bandwidth and sensitivity using classical (black) and practical entangled probe (red) in main text Fig. 3 (c,d). Vertical white dashed line in (c): same as bandwidth using optimal entangled probe (blue) in the main text Fig. 3 (d). Simulation parameters are the same in our experiment.

the optimal and experimental entangled probes enhance the force sensitivity over the classical probes by suppressing the shot noise at large frequency differences.

It is straightforward to derive the SBPs associated with the classical and optimal entangled probes using Eq. (S39):

$$\mathcal{S} \times \mathcal{B}_{3\text{dB}} \approx \frac{\beta \alpha_c}{\omega_{\min} m_{\text{eff}}} \sqrt{\frac{1}{S_{\bar{F}_{\min}\bar{F}_{\min}} S_{Y_0 Y_0}}} \quad \text{(S40)}$$

In the limit of thermal-noise-dominant minimum sensitivity, i.e., $S_{\bar{F}_{\min}\bar{F}_{\min}} \approx S_{\bar{F}_{\text{th}}\bar{F}_{\text{th}}}$, entangled probes improve the SBP by $\sqrt{e^{2r}}$. When the joint imprecision noise is much larger than the thermal noise due to, e.g., the large resonance frequency difference between the two sensors or weak probe power, the entangled probes can enhanced the SBP by a factor of $e^{2r}$. Regardless of the more prominent quantum advantage, homogeneous resonance frequencies are desired in general to maximize the SBP.

## F. Incoherent force sensing

Optomechanical sensors are ideally suited for weak incoherent force sensing. As an example, we consider joint estimation of weak random forces $F_{\text{sig}}^{(i)}$ buried in the thermal and imprecision noise at two optomechanical sensors. We assume the random forces are Gaussian white noise and uncorrelated at two sensors with PSD much less than the force-noise PSD, namely $S_{F^{(i)}F^{(i)}}^{\text{sig}} \ll S_{\bar{F}\bar{F}}$. We choose an unbiased force power estimator $E(t) = 1/t \int_0^t F_{\text{est}}^2(\tau) d\tau$ with

7

mean

$$\langle E(t) \rangle = \left\langle 1/t \int_0^t F_{\text{est}}^2(\tau) d\tau \right\rangle \tag{S41}$$

$$= 1/t \int_0^t \langle \bar{F}_N^2(\tau) \rangle d\tau + 1/t \int_0^t \langle \bar{F}_{\text{sig}}^2(\tau) \rangle d\tau \tag{S42}$$

$$= \delta^2 \bar{F}_N + \delta^2 \bar{F}_{\text{sig}}. \tag{S43}$$

Here, we denote the equivalent joint noise force $\bar{F}_N$ with noise PSDs given in Eq. (S31, S32), subject to the probe power. To resolve weak signals embedded in the noise background, the imprecision of the estimator need be less than the power of signal, i.e., $\delta E(t) < \delta^2 \bar{F}_{\text{sig}}$. As such, we define the equivalent force spectral resolution (EFSR) as

$$\delta F_E(t) = \sqrt{\delta E(t)} \approx \sqrt{\delta E_N(t)} \tag{S44}$$

in the limit of weak signals $\bar{F}_{\text{sig}} \ll \bar{F}_N$, where $E_N = 1/t \int_0^t \bar{F}_N^2(\tau) d\tau$. Over a long sampling time, the EFSR can be well approximated by

$$\delta F_E(t) = \delta \bar{F}_N / (t B_s)^{1/4}, \tag{S45}$$

where $B_s$ is the sampling bandwidth. The EFSR's time dependence of $(t B_s)^{-1/4}$ is consistent with previous reports[4]. In the case of shot-noise-dominant $\bar{F}_N$ due to, e.g., a large frequency difference between the two sensors, entangled probes reduce the integration time in achieving the same force resolution attained by the classical probes. On the contrary, both the classical and entangled probes require a similar integration time to achieve the same force resolution when the force noise $\bar{F}_N$ is overwhelmed by the thermal noise. In this regime, entangled probes enlarge the sensing bandwidth to effectively increase the scanning rate in searches for weak signals with unknown frequencies. Similar behaviours have been observed in squeezed-light-enhanced haloscope in searches for the axion dark matter[5].

## II. SAMPLE DESIGN AND FABRICATION

Fabrication begins by coating a 1.5 um thick photoresist (S1813) on a double-sided 100 nm thick silicon nitride ($Si_3N_4$) on a silicon (Si) wafer. The resist on one side of the wafer is patterned in the shape of a square window using a photolithography system (MLA-150), while the resist on the other side protects the wafer from handling scratches. After developing the exposed area, the pattern is transferred to the silicon nitride using fluorine-based (Ar+SF6) reactive ion etch. After removing the remaining resist, chips are thoroughly cleaned using hydrofluoric acid (HF), water, and Isopropanol (IPA) and mounted on a Teflon holder compatible with strong acid and bases. Then the mount is placed in a potassium hydroxide (KOH) bath at 80 °C for 21 hours to wet etch the silicon in the patterned region and subsequently release the 100 um wide membrane on the other side of the wafer. The released membrane is dried using a gradual dilution process, including iteratively replacing KOH with DI water followed by a 10 min HF dip, IPA, and methanol rinse. Finally, the chips are dried in air and optically bonded on a clean mirror to realize a membrane on a mirror cavity system.

## III. EXPERIMENTAL DETAILS

### A. Squeezed light generation

Squeezed light at 1550 nm is generated by pumping an optical parametrical amplifier (OPA) cavity below threshold. The pump light at 775 nm is phase locked to amplify the seed laser at 1550 nm through a type-0 periodically-poled KTiOPO$_4$ (PPKTP) nonlinear crystal embedded in an optical cavity. The output power of the squeezed light can be effectively tuned by adjusting the seed laser power. All cavity lengths are locked based on the Pound-Drever-Hall technique using 24-MHz sidebands created by phase modulating the 1550-nm pump laser. More details of squeezed-light source are reported in the Supplemental Material of a previous work [6]. We upgrade the solid-state laser in

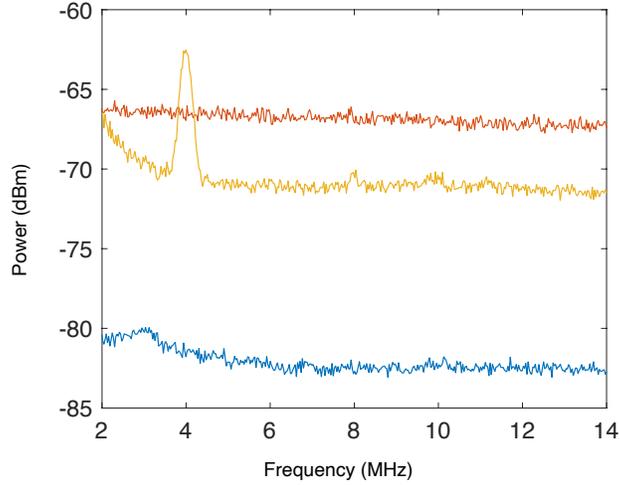

FIG. S2. Spectrum of squeezed light. Yellow: squeezed light. Red: shot noise. Blue: electrical noise. RBW: 300 KHz. VBW: 300 Hz. 4 dB squeezing above 5 MHz. 4 MHz peak is the beating frequency between two phase locking signals.

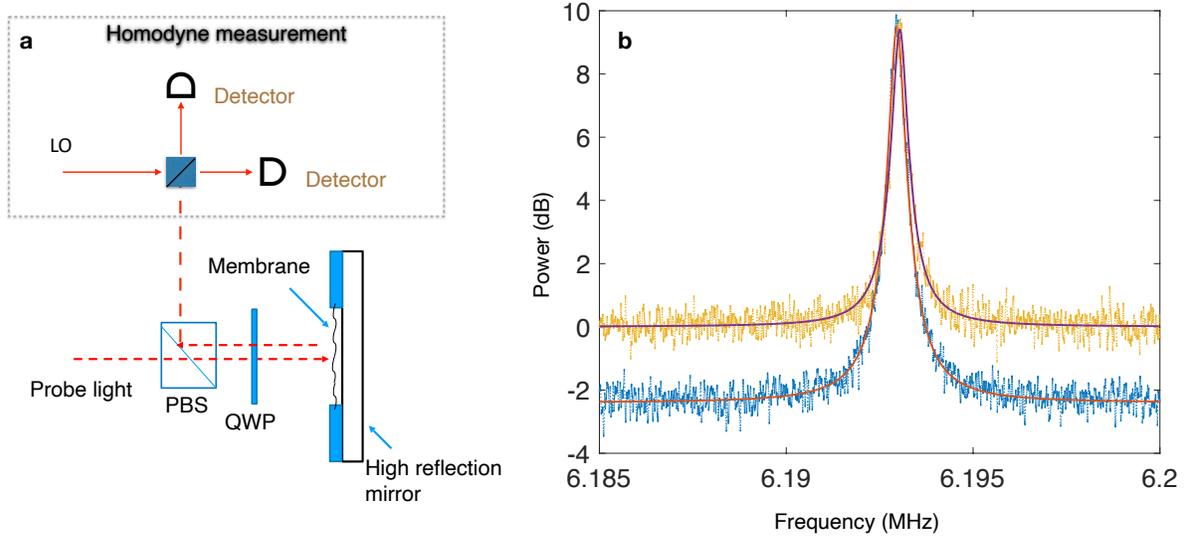

FIG. S3. (a) Optomechanical sensor. QWP: quarter-wave plate; PBS: polarizing beam splitter. LO: local oscillator. (b) Measured displacement noise spectra. Yellow: classical light. Blue: squeezed light. Solid lines: theory fit. Shot noise is normalized to unity. RBW: 20 Hz, VBW: 10 Hz. Traces are averaged over 50 times.

earlier experiments to a low-noise fiber laser (NP Photonics, Power Rock). The squeezing spectrum is plotted in Fig. S2. We observe squeezing below the shot-noise level down to 2 MHz and 4 dB squeezing above 5 MHz. To fully exploit the squeezed light, we address the first higher-order mode of the optomechanical sensor at around 6 MHz, in lieu of its fundamental resonance frequency at around 4 MHz.

9

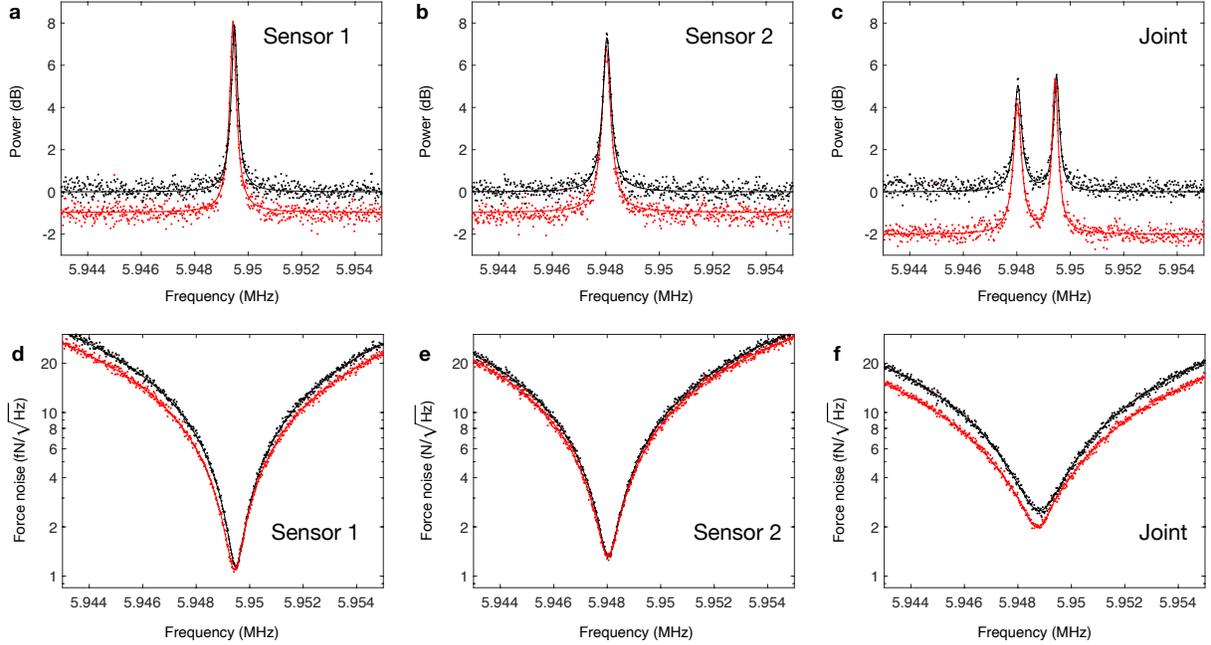

FIG. S4. PSDs and force sensitivities at two sensors with frequency difference 1422 Hz. PSD at (a) sensor 1, (b) sensor 2. (c) Joint PSD replicated from main text Fig. 2(a). Force sensitivity at (d) sensor 1, (e) sensor 2. (f) Joint force sensitivity replicated from the main text Fig. 2(d).

### B. Measurement and calibration of mechanical displacement

Our optomechanical sensor comprises a $100 \times 100$ $\mu m^2$ $Si_3N_4$ membrane with a reflectively $R \approx 11.5\%$ atop a high reflectively ($R > 99.9\%$) mirror, forming a Finesse $\sim 3$ optical cavity, as shown in Fig. S3 (a). The sensors are placed inside a vacuum chamber. To measure the membrane displacement, the probe light is first converted to circular polarization by a QWP and then focused onto the membrane. The optomechanical coupling efficiency is maximized when the optical mode spot is centered near the antinodes of mechanical mode. The displacement is encoded on the phase of reflected probe which is then separated by the same QWP and PBS. The phase quadrature is subsequently measured by a homodyne detector. The propagation loss is measured to be around 18% before interfering with a local oscillator (LO). The interference visibility of homodyne measurement is optimized to 92%. The dc signal of the photocurrent is used to lock the phase between the LO and the probe light. The ac component of photocurrent is sampled by a time-domain spectrum analyzer. The measured PSDs with the classical and squeezed probes are shown in Fig. S3 (b). The thermal peaks for both probes near the resonance frequency due to the Brownian motion of mechanical oscillator almost overlap. The shot noise away from the resonance frequency is suppressed by squeezed light. We observe 2.4 dB squeezing, consistent with the overall loss. The measured PSDs are fitted using Eq. (S13) to extract the optomechanical coupling efficiency $\beta$ and the mechanical susceptibility $\chi(\omega)$. The resonance frequency slightly drifted over the measurements based on classical and squeezed probes. The fitted mechanical linewidths with the classical and squeezed probes are both around 358 Hz, differing less than 1%.

### C. Calibration of individual sensors

We balance the optomechanical coupling efficiencies at the two sensors by slightly displacing the position of the light spot on one membrane. With identical coupling efficiencies, near-optimum entangled probes can be generated via splitting the squeezed light evenly into two arms so that each sensor receives 50% of the squeezed light. The resonance frequency disparity between the fabricated membranes can be as large as 300 kHz. We pick two membranes

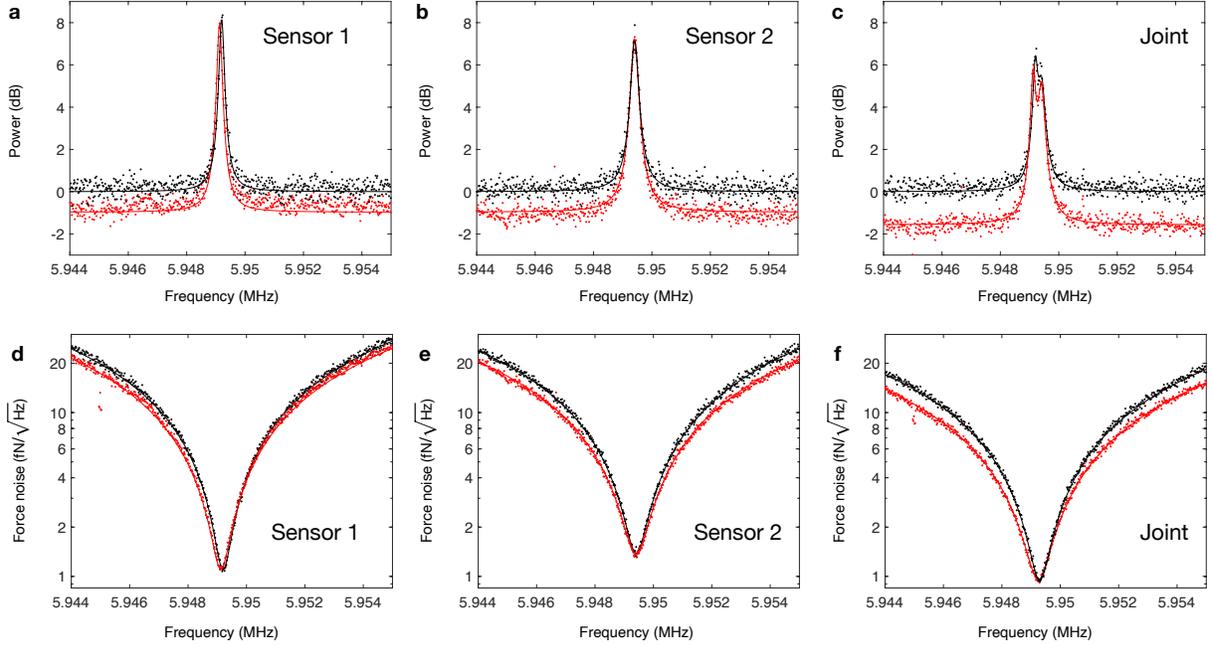

FIG. S5. PSDs and force sensitivities at two sensors with frequency difference 262 Hz. PSD at (a) sensor 1, (b) sensor 2. (c) Joint PSD replicated from main text Fig. 2(a). Force sensitivity at (d) sensor 1, (e) sensor 2. (f) Joint force sensitivity replicated from the main text Fig. 2(d).

with similar resonance frequencies with difference close to the linewidth and then tune the resonance frequency by shining thermal light (Thorlabs OSL2) onto the chip.

The entangled probes are measured by two homodyne detectors. The photocurrents are filtered by high-pass filters (Thorlabs EF513), demodulated by an electrical mixer, and amplified by a low-noise voltage preamplifier (Stanford Research Systems SR560). The data are acquired by an oscilloscope (LeCroy Waverunner 604HD) at 1 MHz sampling rate for 20 seconds and post processed to derive the individual and joint PSDs. An example of joint PSD is presented in Fig. 2(a) of the main text. The associated PSDs at individual sensors are shown in Fig. S4 (a, b), while the joint PSD in (c) reproduced from Fig. 2(a) of the main text is plotted as a comparison. Since each sensor only receives half of the squeezed light, the measurement noise is merely 1 dB below the shot-noise level However, due to the quantum correlation between the measurement noise at the two sensors, the joint noise floor using entangled probes is reduced to 2 dB below the shot-noise level. The magnitude of the two thermal peaks remains unchanged near the two resonance frequencies because the thermal noise at the two sensors is uncorrelated. The force sensitivities of the individual sensors obtained using Eq. (S14) are plotted in Fig. S4 (d,e). In contrast to the shot-noise-dominant joint force sensitivity, the force sensitivity at each individual sensor is limited by the thermal force. The entangled probes only marginally improve the sensing bandwidth. The minimum joint force noise for the classical probes is around 2 times higher than that of an individual sensor, but the bandwidth of the joint force sensing is around 2.8 times larger than that of an individual sensor. The configuration of large frequency difference is suitable for broadband sensing while entangled probes further improve the measurement sensitivity.

On the other hand, when the two resonance frequencies are close, the joint force sensitivity is limited by the thermal noise for both entangled and classical probes, as shown in Fig. S5 (c,f) replicated from Fig. 2(d,e) of the main text. As a comparison, the corresponding PSDs and force sensitivities at each sensor are presented in Fig. S5 (a,b) and (d,e). The joint force sensitivity is improved by $\sqrt{2}$ compared to the force sensitivity of single sensor while entangled probes further enlarges the sensing bandwidth.

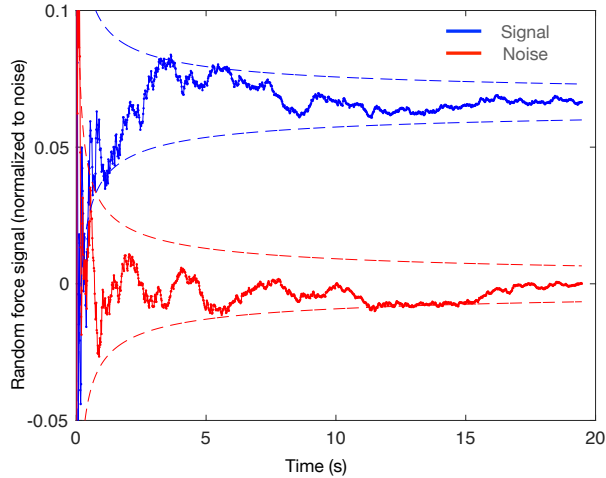

FIG. S6. Contribution percentage of force signal to total force noise $(E(t_i) - \delta^2 \bar{F}_N)/\delta^2 \bar{F}_N$. Red: force noise only with zero mean. Blue: signal force with a displacement. Dashed line: estimation uncertainty of the force power. The measured force power is well confined between the dashed lines.

### D. Radiation pressure force sensing

We are interested in detecting averaged stationary incoherent signal introduced by the radiation pressure force. Two laser beams at 780 nm with randomly modulated amplitude are sent to drive the membranes. The amplitude modulations are uncorrelated such that we ignore the mechanical induced phase shift $\theta_i$. The measured phase quadrature $Y_{\text{out}}^{(i)}(t)$ at two sensors are first converted to the equivalent sampled force $F^{(i)}(t)$ by the mechanical response functions. Then individual sampled force $F^{(i)}$ is filtered by a bandpass filter with a bandwidth obtained from Eq. (S39) during post-processing. The average joint force is given by $\bar{F}_{\text{est}}(t) = (F^{(1)}(t) + F^{(2)}(t))/2$. The estimated force power at different time $t_i$ is $E(t_i) = 1/t_i \int_0^{t_i} \bar{F}_{\text{est}}^2(t)dt$. The normalized signal force power $(E(t_i) - \delta^2 \bar{F}_N)/\delta^2 \bar{F}_N$ is plotted in Fig. S6. The red line shows the noise force power without signal. The mean of noise converges to zero at large integration time. The dashed lines is the statistical uncertainty fitted from Eq. (S45). Turning on the signal force, we observe a constant displacement of the total power, shown as blue line in Fig. S6. The signal can be resolved as long as the displacement is above the statistical uncertainty of the total noise. Here, the applied signal is about only 6.6% of the total noise and is clearly resolvable after about 1 second.

In the limit of weak signals $\delta^2 \bar{F}_{\text{sig}} \ll \delta^2 \bar{F}_N$, the statistical uncertainty of estimator mainly arises from noise. Without loss of generality, we only characterize the noise properties of our optomechanical sensors. To determine the accuracy of the power estimator, we recorded $N = 20$ independent measurement of $\bar{F}_N(t)$. Each of the measured traces gives an estimator of $E^{(n)}(t_i)$ with respect to the acquisition time $t_i$ up to 1 second. The root mean square error of the estimator is calculated by $\delta E(t_i) = \sqrt{\sum_{n=1}^{N} [E^{(n)}(t_i)]^2 / N}$. The EFSR after a $t_i$-second averaging thus reads

$$\delta F_E(t_i) = \sqrt{\delta E(t_i)}. \tag{S46}$$

Two examples of the measured EFSR are presented in Fig. 5 (a,b) of the main text. The fitted solid lines are given by Eq. (S45).

---

[1] A. A. Clerk, M. H. Devoret, S. M. Girvin, F. Marquardt, and R. J. Schoelkopf, Reviews of Modern Physics **82**, 1155 (2010).
[2] M. Aspelmeyer, T. J. Kippenberg, and F. Marquardt, Reviews of Modern Physics **86**, 1391 (2014).
[3] M. Korobko, L. Kleybolte, S. Ast, H. Miao, Y. Chen, and R. Schnabel, Physical Review Letters **118**, 143601 (2017).



\end{document}